\documentclass[journal]{IEEEtran}

\usepackage{cite}
\usepackage{multirow}
\usepackage[cmex10]{amsmath}
\usepackage{amssymb}
\usepackage{array}
\usepackage{url}
\usepackage{ctable}
\usepackage{afterpage}
\usepackage{booktabs}
\usepackage{footnote}
\usepackage{threeparttable}
\usepackage{colortbl}
\usepackage{float}
\usepackage{filecontents}
\usepackage{subfig}
\usepackage[normalem]{ulem}

\begin{document}
\bstctlcite{IEEEexample:BSTcontrol}

\title{Integrated Conductivity-Modulation-Based RF Magnetic-Free Non-Reciprocal Components: Recent Results and Benchmarking}

\author{Negar~Reiskarimian,~\IEEEmembership{Student Member, IEEE,}
        Aravind~Nagulu,~\IEEEmembership{Student Member, IEEE,}
        Tolga~Dinc,~\IEEEmembership{Member, IEEE,}
        and~Harish~Krishnaswamy,~\IEEEmembership{Member, IEEE,}
}

\markboth{Submitted to IEEE Antennas and Wireless Propagation Letters}{}

\maketitle

\begin{abstract}

Achieving non-reciprocity and building non-reciprocal components through spatio-temporal modulation of material properties has attracted a lot of attention in the recent past as an alternative to the more traditional approach of exploiting Faraday rotation in magnetic materials. In this letter, we review recent research on spatio-temporal conductivity-modulation, which enables low-loss, small-footprint, wide-bandwidth and high-power-handling non-reciprocal components operating from radio frequencies (RF) to millimeter-waves (mm-waves) and integrated in a CMOS platform. Four generations of non-reciprocal circulators and circulator-based systems will be reviewed. We will also discuss metrics of performance that are important for wireless applications and standards, and introduce a new antenna (ANT) interface efficiency figure of merit ($\eta_{ANT}$) to enable a fair comparison between various types of antenna interfaces.

\end{abstract}

\section{Introduction}
\label{introduction}

\IEEEPARstart{R}{eciprocity} mandates the same signal transmission profile for waves propagating in opposite directions between two points in space. Various theorems of reciprocity has been formulated over the years, including the work of Green (electrostatics), Rayleigh (dynamic systems), Helmholtz (optics, acoustics) and Lorentz (electromagnetics) \cite{achenbach2003reciprocity}. Lorentz reciprocity is a fundamental physical precept that characterizes the vast majority of electronic and photonic materials, circuits and components.

Non-reciprocal components, such as circulators, isolators, gyrators and non-reciprocal phase-shifters, are critical for various RF and mm-wave applications, including communications, radar, imaging and sensing. Specifically, high-performance circulators can find application as the shared-antenna interface of transceivers for emerging wireless communication paradigms such as full-duplex (FD) \cite{Flexicon_Comm_Mag,FullDuplex_Katti,FullDuplexInvitedPaper_RiceJSAC14} and FD multiple-input multiple-output (MIMO) systems \cite{FD_MIMO_Trans_Comm_2017}, and in frequency-modulated continuous-wave (FMCW) radar. FD aims to instantly double the link capacity in the physical layer by simultaneously transmitting and receiving at the same frequency, as well as providing other benefits in the higher layers \cite{FullDuplex_Katti,FullDuplexInvitedPaper_RiceJSAC14}.

Reciprocity can be violated by breaking one of three necessary conditions - time-invariance, linearity or by exploiting materials with an asymmetric permittivity or permeability tensor. Traditionally, non-reciprocity has been achieved by using magnetic materials such as ferrites, which lose their reciprocity under the application of an external biasing magnetic field. Such magnetic non-reciprocal components are limited in their applicability, since ferrites are incompatible with IC fabrication processes and a permanent magnet is required to induce the magneto-optic Faraday effect \cite{saleh1991fundamentals}. This causes the resulting non-reciprocal element to be bulky, expensive and incompatible with CMOS integration. The oldest alternative to using magnets is to exploit the inherent non-reciprocity of active current/voltage-biased transistors \cite{Tanaka1965_ActiveCirculator,Caloz_MNM,kodera2013magnetless}. Such an approach is compatible with IC fabrication, but is limited by the noise and nonlinearity introduced by the active devices \cite{ActiveCirculator_TMTT2000}. Breaking reciprocity through nonlinearity has also been explored at RF \cite{sounas2018broadband} as well as in the optical domain \cite{fan2012all}. However, the signal-power-dependent performance limits the applicability of this approach. 

In recent years, there has been progress on breaking reciprocity through time-variance, specifically spatio-temporal modulation of material parameters. Such approaches are theoretically linear to the desired signal and noise free. Early approaches have focused on permittivity as the modulated parameter \cite{sounas2017,lira2012electrically,tzuang2014non,Estep_NaturePhys,AustinCirculator_TMTT16, Kord_TMTT2018,NR_UCLA_TMTT14,chamanara2017optical,taravati2017nonreciprocal}. In the optical domain, permittivity is modulated through electro-optic or acousto-optic interactions, but the modulation index is typically extremely weak. In the RF domain, permittivity modulation is achieved using varactors which too exhibit limited modulation index ($C_{max}/C_{min}$ is usually around $2-4$). A weak modulation index directly translates to a large device size over which modulation must be performed. Varactors, and permittivity modulation in general, also exhibit a trade-off between modulation index and loss, particularly as the operating frequency is increased. Consequently, these efforts have resulted in designs that exhibit a trade-off between loss, size, bandwidth and linearity. 

In the circuits community, time-modulated systems are commonly called linear periodically-time-varying (LPTV) circuits. Early reports on LPTV circuits date back to the 1940s \cite{1947_commutation_paper}. In the following two decades, a class of LPTV circuits and systems also referred to as commutated networks attracted a lot of attention, and theoretical \cite{1953_commutation_paper,1969_commutated_paper} and practical aspects \cite{Franks_ISSCC_1960} of their implementation were explored. More recently, commutated networks have emerged once again as a hot topic of research, primarily due to the ability to realize tunable RF high-quality-factor filters (the so-called “N-path filters”) in CMOS for the first time \cite{Npath_Ghaffari,Molnar_mixer_first,NR_RFIC2015}. Recently, we found that commutated networks have a rich set of unique properties that go beyond high-Q filtering, including the ability to realize non-reciprocity \cite{NRK_NatComm16,NRK_JSSC2017,NRK_ISSCC2017,TD_NatComm17,Tolga_Circ_JSSC,Aravind_RFIC2018}.

\begin{figure*}[!t]
	\centering
	\includegraphics[keepaspectratio,width=1\linewidth]{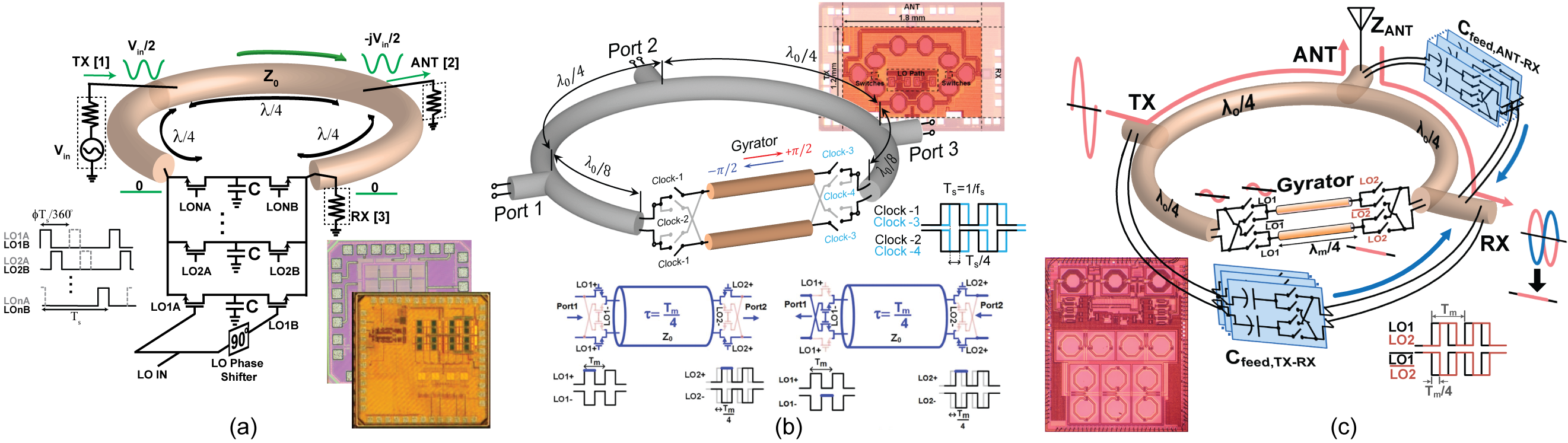}
	\caption{Conceptual diagrams of our conductivity-modulation-based circulators: (a) RF N-path-filter-based circulator and circulator-receiver architecture employing a linearity-enhancement technique for transmitter side excitations \cite{NRK_JSSC2017,NRK_ISSCC2017}. (b) Millimeter-wave switched-transmission-line circulator \cite{TD_NatComm17,Tolga_Circ_JSSC}. (c) RF switched-transmission-line circulator with $>$1W power-handling capability and inductor-free antenna balancing \cite{Aravind_RFIC2018}.}
	\label{fig:review_fig}
\end{figure*}

In this letter, we will cover our recent research on using commutated circuits to break reciprocity and build high-performance passive circulators operating from RF to mm-wave frequencies in CMOS. \emph{Our switch-based commutated circuits may be fundamentally understood as performing spatio-temporal conductance modulation, and leverage the fact that conductivity is a variable material property that is unique to semiconductors. Conductivity in semiconductors can be modulated over a wide range (CMOS switch ON/OFF conductance ratios can be as high as $10^3-10^5$) relative to permittivity.} In Section \ref{concept}, we will describe the fundamental physical principles, as well as four generations of CMOS circulators and circulator-based wireless communication systems \cite{NRK_JSSC2017,NRK_ISSCC2017,TD_NatComm17,Tolga_Circ_JSSC,Aravind_RFIC2018} targeting emerging full-duplex and 5G mm-wave applications. We will also cover architectural concepts that improve the linearity and isolation. In Section \ref{AntennaFoM}, we discuss various metrics of performance that are critical for wireless applications, and introduce a new antenna interface efficiency figure of merit ($\eta_{ANT}$) that enables a fair comparison between various reciprocal and non-reciprocal shared-antenna interfaces. Finally, Section \ref{conclusion} concludes the paper.

\section{Conductivity-Modulation-Based Non-Reciprocity}
\label{concept}

\subsection{N-Path-Filter-Based Circulator and Circulator-Receiver Architecture}

N-path filters enable the implementation of reconfigurable, high-$Q$ filters at RF in nanoscale CMOS IC technology \cite{Npath_Ghaffari}, and are a class of LPTV networks where the signal is periodically commutated through a bank of capacitors. We have found that applying a relative phase shift to the non-overlapping clocks driving the input and output switch sets of a two-port N-path filter imparts a non-reciprocal phase-shift to the signals traveling in the forward and reverse directions since they see a different ordering of the phase-shifted switches \cite{NRJZ_TCAS2}. \emph{Essentially, the two-port N-path filter with a clock phase shift of 90$^\circ$ realizes an electrically-infinitesimal gyrator.}

\begin{figure*}[!t]
	\centering
	\includegraphics[keepaspectratio,width=0.9\linewidth]{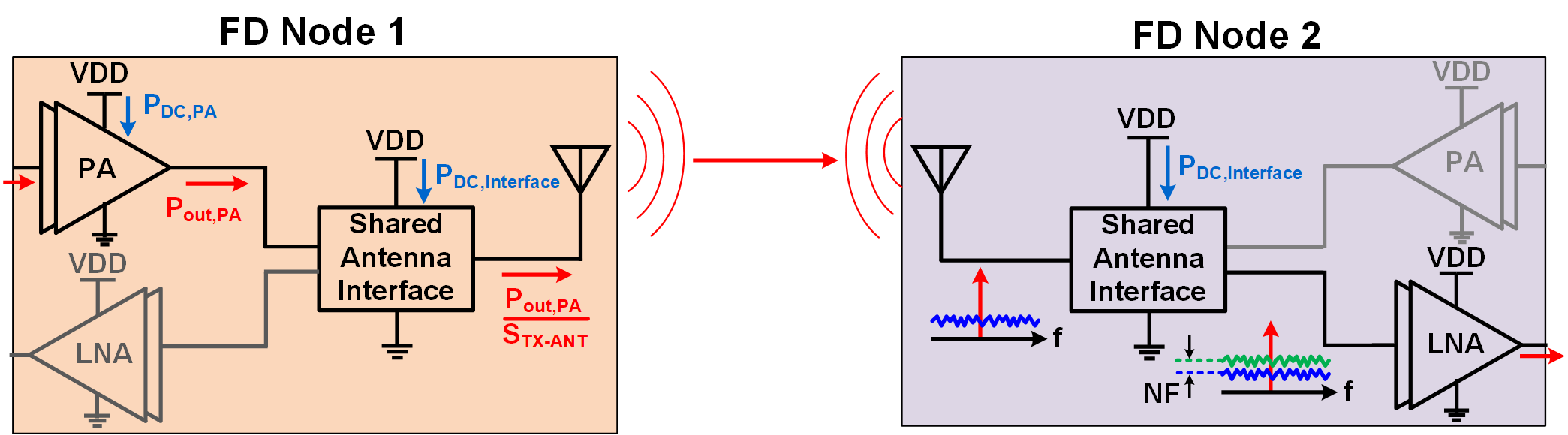}
	\caption{FD wireless communication link using shared-antenna interface.}
	\label{fig:FOM_Link}
\end{figure*}

To convert phase non-reciprocity to non-reciprocal wave propagation, an N-path-filter with $\pm 90^\circ$ phase-shift is placed inside a transmission line loop with a length of 3$\lambda$/4 (Fig. \ref{fig:review_fig}(a)). The combination of the non-reciprocal phase shift of the N-path filter with the reciprocal phase shift of the transmission line results in  unidirectional wave propagation ($-270^\circ-90^\circ$=$-360^\circ$), because the boundary condition for wave propagation in the reverse direction can not be satisfied ($-270^\circ+90^\circ$=$-180^\circ$). Additionally, a three-port circulator can be realized by placing ports anywhere along the loop as long as they maintain a $\lambda$/4 circumferential distance between them. \textit{Interestingly, maximum linearity with respect to the transmitter (TX) port is achieved if the receiver (RX) port is placed adjacent to the N-path filter, since the inherent TX-RX isolation suppresses the voltage swing on either side of the N-path filter, enhancing its linearity.} Table \ref{tab:comparison} lists the performance metrics of a 750MHz 65nm CMOS N-path-filter-based circulator, \emph{which is the first CMOS passive circulator to be implemented} \cite{NRK_NatComm16,NRK_JSSC2017}. Furthermore, the N-path filter in the circulator can be repurposed as a down-converting mixer, directly providing the baseband received signals on the N-path filter capacitors. More details about this work can be find in \cite{NRK_ISSCC2017}.

\subsection{Wideband Millimeter-Wave Swithced-Transmission-Line-Based Non-Reciprocal Circulator}

Inspired by the N-path-filter-based RF circulator discussed in the previous subsection, we proposed a  generalized conductivity modulation concept using switched transmission lines \cite{TD_NatComm17,Tolga_Circ_JSSC}. The concept, shown in Fig. \ref{fig:review_fig}(b), consists of two sets of differential mixer-quad switches on either end of a differential transmission-line delay (replacing the commutated capacitors). The switches are clocked at a modulation frequency $\omega_m$, and the delay of the line is equal to one quarter of the modulation period ($T_m /4$). The switches are clocked with 50\% duty-cycle square-wave clocks with a relative phase shift of $T_m/4$. As a result of this, waves traveling from left to right experience the transmission-line delay with no sign flips in both halves of the clock period. On the other hand, waves traveling from right to left experience the transmission-line delay along with one sign flip. In other words, transmission in both directions is perfectly lossless, but there is an infinitely broadband $180^\circ$ non-reciprocal phase difference. \emph{In other words, the structure realizes an infinitely broadband gyrator.} Similar ideas related to switched transmission lines to realize broadband non-reciprocity have been explored in \cite{biedka2017ultra,krol2017non}.

The key aspect of this architecture is that the infinite bandwidth of the gyrator implies that the signal frequency and the modulation frequency are completely decoupled. An arbitrarily low modulation frequency can be used, with the only restriction being an associated increase in the transmission line length, and hence, loss. This benefit was exploited to realize a mm-wave (25GHz) fully-integrated passive circulator in 45nm SOI CMOS, shown in Fig. \ref{fig:review_fig}(b) \cite{TD_NatComm17,Tolga_Circ_JSSC}, by once again placing a $3\lambda /4$ transmission line around the gyrator. Modulation was performed at 1/3rd of the operating frequency (8.33GHz). The lowering of the modulation frequency, as well as the need for 50\% duty-cycle clocks, as opposed to numerous low-duty-cycle clocks as in N-path filters, is critical as clocking switches at mm-wave frequencies is impractical in current CMOS IC technology. The infinitely-broadband phase non-reciprocity also implies broader bandwidth of operation for the resultant circulator. A summary of the performance of the 25GHz prototype can be found in Table \ref{tab:comparison}.

\subsection{Highly Linear RF Non-Reciprocal Circulator with Loss-Free, Inductor-Free Antenna Balancing}

Fig. \ref{fig:review_fig}(c) shows the architecture of a highly-linear RF circulator, which uses the feature of lowering the modulation frequency to enhance the power handling and linearity of the circulator at 1GHz operating frequency. The switches were modulated at 333MHz for 1GHz operation, and such a low modulation frequency enables the usage of the thick-oxide devices in 180nm SOI CMOS technology to boost power handling. The measured in-band TX-ANT/ANT-RX input third-order intercept points (IIP3) are +50dBm and +36.9dBm respectively. TX-ANT input 1dB compression point (P1dB) is $>$+30.7dBm (limited by the measurement setup), at which point the compression is only 0.66dB, while the ANT-RX input P1dB is +21dBm. This implementation notably improves linearity and power handling by 10-100$\times$ when compared with our prior CMOS non-reciprocal circulators~\cite{Aravind_RFIC2018}.

The TX-RX isolation of all shared-antenna interfaces is limited by the matching of the antenna port, which necessitates an antenna tuning mechanism. Traditionally, magnetic circulators are followed by antenna tuners. \textit{In this work, antenna balancing is achieved by implementing digitally programmable feed capacitor banks between TX-RX and ANT-RX ports.} Due to the $-90^\circ$ phase difference between the TX and ANT ports and the programmable nature of the feed capacitors, this produces tunable orthogonal currents which can cancel the leakage at the RX port produced by an antenna mismatch (Fig. \ref{fig:review_fig}(c)). The fully-differential architecture enables a free sign flip in the feed capacitor banks, and therefore allows complete VSWR coverage without using any inductive or resistive elements in the feed circuitry. Aggressive device stacking is employed in the static switches used in these feed capacitor banks so that they do not limit linearity and power handling. The ability to compensate a VSWR of up to 1.85 (and beyond) was demonstrated in this work. A summary of the performance of the prototype can be found in Table \ref{tab:comparison}.

\section{Antenna Interface Efficiency FoM}
\label{AntennaFoM}

\begin{table*}[!t]
	\caption{Comparison Table}
	\vspace{-0.15cm}
	\scriptsize
	\centering
	\renewcommand{\arraystretch}{1.4}
	\renewcommand*{\arraystretch}{1.3}
	
	\begin{tabular}{|c|c|c|c|c|c|}\hline
		& {\bfseries \cite{NRK_JSSC2017}}& {\bfseries \cite{TD_NatComm17}}& {\bfseries \cite{imec_EBD_ISSCC2015}}& {\bfseries \cite{Aravind_RFIC2018}}\\ 
		\hline
		Architecture & N-path-filter circulator & \parbox[c]{36mm}{Switched-t-line mm-wave circulator} & Electrical-balance duplexer & Highly-linear RF circulator  \\ \hline
		Technology & 65nm CMOS & 45nm SOI CMOS & 180nm SOI CMOS&  180nm SOI CMOS  \\ \hline
		Frequency & 0.61 - 0.975GHz & 25GHz & 1.9 - 2.2GHz &  0.86 - 1.08GHz \\ \hline
		TX-ANT/ANT-RX Trans. & -1.7/-1.7dB & -3.3/-3.2dB& -3.7/-3.9dB&  -2.1/-2.9dB \\ \hline	
		Isolation BW & \parbox[t]{10mm}{1.9$\%$($>$25dB), 0.33$\%$($>$40dB)} & 18$\%$ ($>$18.5dB) $^2$ & $~$15$\%$ ($>$40dB)& \parbox[t]{10mm}{17$\%$($>$25dB), 3.1$\%$($>$40dB)}\\ \hline	
		Z$_{ANT}$ impedance & 50$\Omega$ & 50$\Omega$& 1.5:1 VSWR&  1.85:1 VSWR \\ \hline	
		ANT-RX NF & 4.3dB $^3$& 3.3 - 4.4dB& 3.9dB& 3.1dB \\ \hline
		TX-ANT IIP3/P1dB & +27.5dBm/ N/R & +20.1dBm/$>$+21.5dBm& +70dBm/$>$+27dBm&+50.025dBm/$>$+30.66dBm \\ \hline	
		ANT-RX IIP3/P1dB & +8.7dBm/ N/R & +19.9dBm/$>$+21dBm& +72dBm/$>$+27dBm& +36.9dBm/+21.01dBm \\ \hline
		TX-induced RX P1dB & N/R & N/R & N/R & +21.3dBm\\ \hline
		Power Consumption & 59$\,$mW & 78.4mW& 0mW& 170mW \\ \hline		
		Chip Area & $\lambda$$^{2}/$6400 or 25mm$^2$ $^4$ & $\lambda$$^{2}/$66 or 2.16mm$^2$& $\lambda$$^{2}/$13000 or 1.74mm$^2$& $\lambda$$^{2}/$5500 or 16.5mm$^2$ \\ \hline 
		$\eta_{ANT}$ $^5$ & 7.2$\%$ & $>$14.9$\%$& 13.8$\%$& $>$23.1$\%$ \\ \hline
	\end{tabular}
	\begin{tablenotes}
		\item[1]	$^1$ $_\textrm{{w/o LNA}}$ - Results from post layout simulation and $_\textrm{{w/ LNA}}$ - Results from measurements , $^2$ Limited by mmWave setup, $^3$Includes 2.3dB degradation due to LO phase noise, $^4$Includes SMD inductors on PCB, $^5$$\eta_{PA} = 30\%$ is used and P1dB is assumed to limit TX power handling. 
		
	\end{tablenotes}
	\label{tab:comparison}
	
\end{table*}

An ideal antenna interface should have no loss, extremely high power handling and no additional power consumption. Conventional three-port reciprocal antenna interfaces, such as hybrids or electrical-balance duplexers (EBDs) \cite{EB_Duplexer_JSSC13,imec_EBD_ISSCC2015,EB_Duplexer_Entesari_TMTT2016,EB_Duplexer_Larson_TMTT2016}, have high power handling but exhibit a fundamental 3dB loss, with an additional $\approx$1dB loss due to implementation issues such as finite metal resistivity and substrate loss. LPTV circulators have lower loss than reciprocal interfaces, but only recently have been achieving comparable power handling and require additional power consumption. Hence, to enable a fair comparison between various shared-antenna interfaces, an FoM called \emph{antenna interface efficiency ($\eta_{ANT}$)} is introduced. It calculates the degradation imparted to the transmitter efficiency by the antenna interface by assuming a certain baseline efficiency.  

\subsubsection{Metrics for Power Handling}

Before introducing the new FoM, it is important to discuss the different metrics that determine the TX-port power handling of an ANT interface. The most common metrics are TX-port input P1dB and IIP3. Typically, communication systems operate with P1dB as the upper limit of the transmitted signal power. IIP3 captures the in-band intermodulation distortion produced on the input signal by the nonlinearity of the interface. This distortion is mainly relevant as it leaks into adjacent channels (a phenomenon commonly called \emph{spectral regrowth}), potentially desensitizing other users. Typical wireless standards mandate a certain adjacent-channel leakage ratio (ACLR). Loosely, operating 20dB below the IIP3 level keeps the leakage into adjacent channels 40dB below the main signal power.

Since non-reciprocal circulators are intended to be used in full-duplex applications, new metrics of power handling arise from the need to receive a desired signal while transmitting a substantial amount of power. \emph{TX-induced RX 1dB compression} is defined as the TX power at which a weak desired signal traveling from the ANT port to the RX port is compressed by 1dB. Based on the results shown in Table \ref{tab:comparison}, this metric is more stringent than P1dB or IIP3. \emph{TX-induced RX NF} refers to the noise figure measured for ANT-to-RX transmission while a powerful transmit signal is fed to the TX port. This metric would include noise generated by reciprocal mixing effects between the TX signal and the modulation signal phase noise. This is a challenging metric to measure, and requires further investigation. Finally, frequency-division duplex (FDD) standards today require as high as +70dBm TX-ANT IIP3 from duplexer filters and EBDs due to a standard-specific test related to the cross-modulation between the TX signal and a so-called full-duplex-spaced (FDS) jammer that would produce spur signals in the RX band.

\subsubsection{Antenna Interface Efficiency}

Consider a half duplex link with no shared-antenna interface, where the power amplifier is directly connected to the antenna. For this scenario, the transmitter efficiency will be equal to the efficiency of the power amplifier ($\eta_{PA}$). A full-duplex link is shown in Fig. \ref{fig:FOM_Link}. In the transmitter of FD node 1, the PA's DC power can be calculated as $P_{DC,PA}=P_{out,PA}/\eta_{PA}$, where $\eta_{PA}$ is the drain efficiency of the PA and $P_{out,PA}$ is the PA output power limited by whichever power handling mechanism discussed earlier dominates. Due to TX-ANT loss of the antenna interface, the output power transmitted would be $P_{out,PA}/(S_{TX-ANT})$. This signal reaches the receiver in FD node 2, where the signal-to-noise ratio is degraded by the ANT-RX noise figure ($NF$) of the antenna interface. Hence, the effective transmitted power of this FD link can be written as $P_{out,PA}/(S_{TX-ANT}\times~NF)$. On the other hand, the DC power consumption of the transmitter is increased to ($P_{DC,PA}+P_{DC,interface}$). The efficiency of the shared-antenna interface, $\eta_{ANT}$, can be expressed as the degradation in transmitter efficiency of this FD link compared to transmitter efficiency in the half-duplex case: 

\begin{equation}
	\eta_{ANT} = \frac{P_{out,PA}/(S_{TX-ANT}\times~NF)}{P_{DC,PA}+P_{DC,interface}}\times\frac{1}{\eta_{PA}}\times 100\%.
	\label{eq:AIE1}
\end{equation}

\emph{In scenarios where power handling is limited by P1dB}, $P_{out,PA}=P_{1dB}$ and at this power level, the TX-ANT loss compresses by 1dB so that $S_{TX-ANT}=~S_{21}\times~1.26$, where $S_{21}$ is the small-signal TX-ANT loss. Hence (\ref{eq:AIE1}) can be modified as

\begin{equation}
	\eta_{ANT} = \frac{P_{1dB}/(S_{21}\times 1.26\times NF)}{(P_{DC,interface}+P_{1dB}/\eta_{PA})}\times\frac{1}{\eta_{PA}}\times 100\%.
	\label{eq:AIE2}
\end{equation}

For an ideal antenna interface, $S_{21}$=0dB, $NF$=0dB, $P_{DC,interface}=$0mW, and $P_{1dB}\to \infty$. Hence, $\eta_{ANT}=100\%$. For an ideal reciprocal interface, such as an ideal EBD, $S_{21}=NF=$3dB, $P_{DC,interface}=$0mW, and $P_{1dB}\to \infty$. From eq.~\ref{eq:AIE1}, the efficiency of an ideal reciprocal interface is $25\%$. In practice, due to the implementation losses, the efficiency of reciprocal interfaces is $<25\%$. For instance, the state-of-the-art EBD in \cite{imec_EBD_ISSCC2015} achieves 13.8\%. For the first time, our work in \cite{Aravind_RFIC2018}, achieves a superior $\eta_{ANT}$ when compared with electrical balance duplexers due to the low-loss and high power handling.

\section{Conclusion}
\label{conclusion}
In this letter, we covered recent advances in conductivity-modulation-based integrated CMOS non-reciprocal components. We also discussed metrics of performance and a new figure of merit that enables a fair comparison across a variety of antenna interfaces. Topics for future research include understanding and mitigating the impact of modulation signal phase noise on the performance of LPTV non-reciprocal components, and exploring how synergies with the acoustic and optic domains can be exploited.

\bibliographystyle{IEEEtran}

\bibliography{reffinal.bib}

\end{document}